\documentclass{nature}
\usepackage{graphicx}
\makeatletter
\let\saved@includegraphics\includegraphics
\AtBeginDocument{\let\includegraphics\saved@includegraphics}
\renewenvironment*{figure}{\@float{figure}}{\end@float}
\makeatother
\usepackage{epstopdf}
\usepackage{dcolumn}
\usepackage{bm}
\usepackage{amssymb}
\usepackage{color}
\usepackage{amsmath}
\usepackage{soul}
\usepackage{float}
\usepackage[normalem]{ulem}

\usepackage{cancel}

\newcommand{\rr}{\mathbf{r}}
\newcommand{\kk}{\mathbf{k}}
\usepackage[
colorlinks=true]
{hyperref}

\hypersetup{colorlinks,
citecolor=blue,
linkcolor=blue,
urlcolor=blue}

\title{Polarization of chirality}

\author{David Ayuso$^{1*}$,
        Andres Ordonez$^{1,2*}$,
        Piero Decleva$^3$,
        Misha Ivanov$^{1,4,5}$,
        Olga Smirnova$^{1,2}$}

\begin{document}
\maketitle

\begin{affiliations}
\item Max-Born-Institut, Berlin, Germany
\item Technische Universit\"at Berlin, Berlin, Germany
\item Dipartimento di Scienze Chimiche e Farmaceutiche, Universit\`a degli Studi di Trieste, Trieste, Italy
\item Institute f\"ur Physik, Humboldt-Universit\"at zu Berlin, Berlin, Germany
\item Department of Physics, Imperial College London, London, UK
\newline *These authors contributed equally to this work
\end{affiliations}

\begin{abstract}

It has been long recognized that the spatial polarization of the electronic clouds in molecules, and the spatial arrangements of atoms into chiral  molecular structures, play crucial roles in physics, chemistry and biology. However, these two fundamental concepts -- chirality \cite{Blaser2013,book_ComprehensiveChiropticalSpectroscopy} and polarization\cite{slater2012electromagnetism} -- have remained unrelated so far.
This work connects them by introducing and exploring the concept of polarization of chirality. We show that, like charge, chirality, or handedness, can be polarized, and that such polarization leads to fundamental consequences, demonstrated here using light. First, we analyze how chirality dipoles and higher-order chirality multipoles manifest in experimental observables. Next, we show how to create chirality-polarized optical fields of alternating handedness in space. Despite being achiral, these racemic space-time light structures interact differently with chiral matter of opposite handedness, and the chirality dipole of light controls and quantifies the strength of the enantio-sensitive response. Using nonlinear interactions, we can make a medium of randomly oriented chiral molecules emit light 
to the left, or to the right, depending on the molecular handedness and on the chirality dipole of light.
The chiral dichroism in emission direction reaches its highest possible value of 200$\%$.
Our work opens the field of chirality polarization shaping of light and new opportunities for efficient chiral discrimination and control of chiral and chirality-polarized light and matter on ultrafast time scales.

\end{abstract}

\begin{figure}[H]
\centering
\includegraphics[width=\linewidth, keepaspectratio=true]{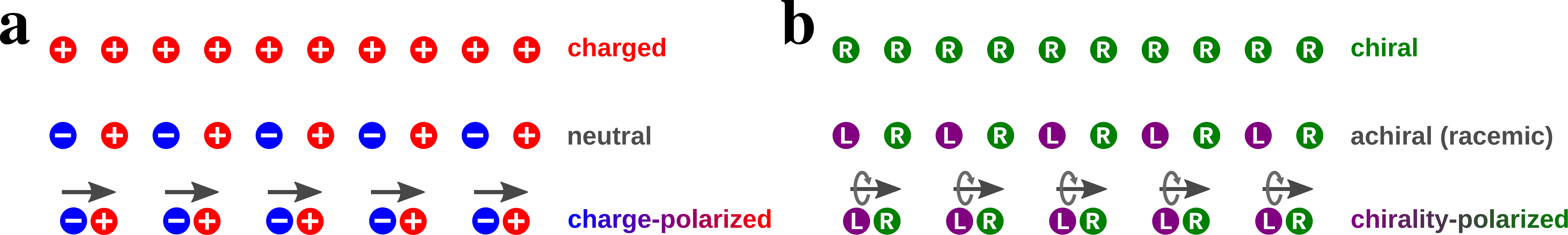}
\caption{\textbf{Polarization of charge versus polarization of chirality.}
\textbf{a,} Mono-dimensional (1D) arrangement of charged units that is: charged and unpolarized, neutral and unpolarized, and neutral and polarized.
\textbf{b,} 1D arrangement of chiral units that is: chiral and unpolarized, achiral and unpolarized, and achiral and polarized.
}
\label{fig_scheme}
\end{figure}

Chirality, or handedness, is a ubiquitous property of light and matter characterized by an unusual type of symmetry -- mirror symmetry. Mirror reflection transforms a chiral object into its opposite counterpart, with our left and right hands being a typical example. 
These mirror twins are called enantiomers, and symmetry dictates that they must behave identically unless interacting with another chiral object. Handedness changes sign upon reflection: two enantiomers are characterized by the handedness of opposite sign. In this sense, handedness is similar to charge: just like a collection of charges can be positive, negative or neutral, a collection of chiral objects can have positive, negative, or zero total handedness (the latter called racemic mixtures). 
Chirality is of tremendous importance in nature and distinguishing molecular enantiomers is vital, which has stimulated a major recent research effort\cite{Nahon2015JESRP,Ritchie1976PRA,Powis2000JCP,Bowering2001PRL,Fischer2005,Belkin2000PRL,Belkin2001PRL, Lux2012Angewandte,Patterson2013Nat, Janssen2014PCCP, Lux2015ChemPhysChem,Comby2016JPCL,yachmenev2016detecting,Eibenberger2017PRL,Beaulieu2018NatPhys,owens2018climbing,yachmenev2019field}. Just as important is the role of spatial polarization of electronic clouds in molecules, from controlling how fatty acids dissolve in water to giving specificity to the biological activity of enzymes\cite{Sakaki2020}. Polarization of charge characterizes overall neutral but non-uniform spatial charge distributions and  their interaction with other charges. 
Following our parallel between distribution of charge and distribution of handedness (see Fig. \ref{fig_scheme}), one would expect \emph{polarization of chirality} to play an important role in the interaction between extended chiral objects displaying a position dependent chirality. Is this expectation correct? 
Can we engineer such structures? How effective is their interaction with chiral media and how can it be characterized? Can polarization of chirality be detected in  an experiment? Here we give positive answers to these fundamental questions.

Let us begin with the simple case depicted in Fig. \ref{fig_scheme}a: a one-dimensional arrangement of alternating positive and negative charges $\pm q$. 
When the particles are uniformly distributed, the medium is not polarized. 
It becomes polarized as we modify their positions, creating dipoles $\mathbf{d}_e=q\mathbf{r}_{0}$, where
$\mathbf{r}_{0}$ is the vector connecting the nearby negative and positive charges.
Consider now a similar racemic distribution of chiral units of alternating handedness, Fig. \ref{fig_scheme}b.
These units can be chiral particles or chiral light fields.
Just like the neutral medium of charged particles, this racemic distribution is  unpolarized if the distances between consecutive chiral units are the same. 
If we modify them, e.g. by shifting the right-handed units to the left, we create \emph{dipoles of chirality}, and the medium acquires \emph{polarization of chirality}.
We could define the dipole of chirality as $\mathbf{d}_{c}=h\mathbf{r}_{0}$,
where  $\mathbf{r}_{0}$ is the vector connecting the nearby left and right-handed units and $h=h_R=-h_L$ is the handedness of  a single unit.

However, the definition of the unit cell in Fig. \ref{fig_scheme}a, together with the polarization of the unit cell, is not unique, a well-known problem encountered in solid state physics\cite{Spaldin2012}. Therefore, the concept of \emph{chirality dipole} should rely on specific observables.
In this context, consider the interaction of chiral matter with chiral light.
Traditionally, chiro-optical effects have relied on the interaction  with both the electric and magnetic field components of a circularly polarized light wave.
However, the magnetic interaction occurs beyond the electric-dipole approximation and is often very weak. This limitation can be bypassed in a particularly straightforward manner by using a pair of non-collinear laser beams. These allow us
to engineer propagating optical fields with three orthogonal polarization components\cite{Hickstein2015NatPhot,Pisanty2018NJP,Ayuso2019NatPhot,Neufeld2019PRX} and create synthetic chiral light\cite{Ayuso2019NatPhot}. Such light is chiral already in the electric-dipole approximation: the electric field of light draws a chiral Lissajous figure in every point in space\cite{Ayuso2019NatPhot}. The chiro-optical response to this light does not rely on magnetic interactions and is orders of magnitude stronger. While synthetic chiral light is chiral in every point in space,  its handedness can be distributed in space as desired. Thus, it is ideally suited to model distributed handedness and test the concept of \emph{polarization of chirality}. 

The enantio-sensitive  optical response to chiral light originates from the interference of two contributions to the light-induced polarization $\mathbf{P}_{\omega}=\mathbf{P}_{\omega}^{ACH}+\mathbf{P}_{\omega}^{CH}$ at a given frequency $\omega$, one of them  not sensitive to chirality ($\mathbf{P}_{\omega}^{ACH}$), and the other unique to chiral matter ($\mathbf{P}_{\omega}^{CH}$), which is out of phase in media of opposite handedness.
The microscopic intensity  $|\mathbf{P}_{\omega}|^2$  is sensitive to the interference of  $\mathbf{P}_{\omega}^{CH}(\rr)$ and  $\mathbf{P}_{\omega}^{ACH}(\rr)$ at the same point $\rr$ through the  term $\mathbf{P}_{\omega}^{ACH*}(\rr)\cdot\mathbf{P}_{\omega}^{CH}(\rr)+\mathrm{c.c.}$ This interference encodes the distributed handedness of light or matter (or both). That is, the near-field response records the distributed handedness locally.  In contrast, the far-field signal provides access to the interference of the chiral and achiral contributions from the whole interaction region and is sensitive to spatial correlations of $\mathbf{P}_{\omega}^{ACH}(\rr)$ and  $\mathbf{P}_{\omega}^{CH}(\rr+\rr')$. The far-field signal maps the distribution of handedness on observables such as the enantio-sensitive direction of light emission and the enantio-sensitive shape of the emission pattern on the screen. These observables are simply the different moments of the enantio-sensitive component  of the  intensity distribution  in the reciprocal space, which is proportional to the real part of
\begin{equation}
\label{eq_Gink-space}
    \tilde{G}_{\omega}(\kk)=\tilde{ \mathbf{P}}_{\omega}^{ACH*}(\kk)\cdot\tilde{\mathbf{P}}_{\omega}^{CH}(\kk).
\end{equation}
(The subscript $\omega$ indicates that we consider far-field signal centered at frequency $\omega$ with bandwidth $\Delta \omega\ll \omega$, hence $\Delta k\ll k$; we  omit this subscript below for brevity.) 
For example, the enantio-sensitive contribution to the total intensity and the average direction of emission are given by the zero and first moments of the distribution, respectively: 
\begin{eqnarray}
    \left<\Delta I_{\omega}\right> \propto\int d^3 k \tilde{G}(\kk) + \text{c.c.},
    \ \ \ \ \left<k_i\right> =\int d^3 k k_i\tilde{G}(\kk) + \text{c.c.}
\label{eq_k_i}
\end{eqnarray}
The enantio-sensitive shape of the emission pattern on the screen is given by the higher moments:
\begin{eqnarray}
    \left<k_{i,j...}\right> &=\int d^3 k k_ik_j...\tilde{G}(\kk) + \text{c.c.},
    \ \ \
\label{eq_k_ijq}
\end{eqnarray}
Eqs. (\ref{eq_k_i}-\ref{eq_k_ijq}) describe the multipoles of the enantio-sensitive intensity distribution in $k$-space.
The different moments of $\tilde{G}(\kk)$ reflect the fact that handedness can have complex distributions both in coordinate and reciprocal space.

If the handedness of matter is distributed uniformly, then  $\tilde{G}(\kk)$ reflects the distributed handedness of light $\tilde{h}(\kk)$, $\tilde{G}(\kk)\propto \tilde{h}(\kk)$.
The enantio-sensitive contribution to light intensity, the direction of light emission, the shape of the light spot on a screen will then encode zero, first, and higher order moments of $\tilde{h}(\kk)$:
\begin{eqnarray}
    h_0=\int d^3 k \tilde{h}(\kk),\ \ \
     \tilde{\mathbf{h}}=\int d^3 k \mathbf{k} \tilde{h}(\kk),\ \ \
      \tilde{h}_{\{ij...\}}=\int d^3 k k_ik_j... \tilde{h}(\kk).
\label{zilch}
\end{eqnarray}
The enantio-sensitive contribution to the total intensity is only non-zero if  light's handedness is non-zero on average, $h_0\neq 0$. But even if $h_0=0$, enantio-sensitive effects remain. For example, if light is racemic, $h_0=0$, but chirality polarized, i.e. $\tilde{\mathbf{h}}\ne0$, we will see enantio-sensitive  deflection in the  nonlinear optical response. 
In general, the distributed handedness of racemic objects manifests itself in an entire array of tensorial enantio-sensitive observables. Their measurement requires acquisition of N-dimensional data sets, where N is the rank of the corresponding tensor. 

\begin{figure}[H]
\centering
\includegraphics[width=\linewidth, keepaspectratio=true]{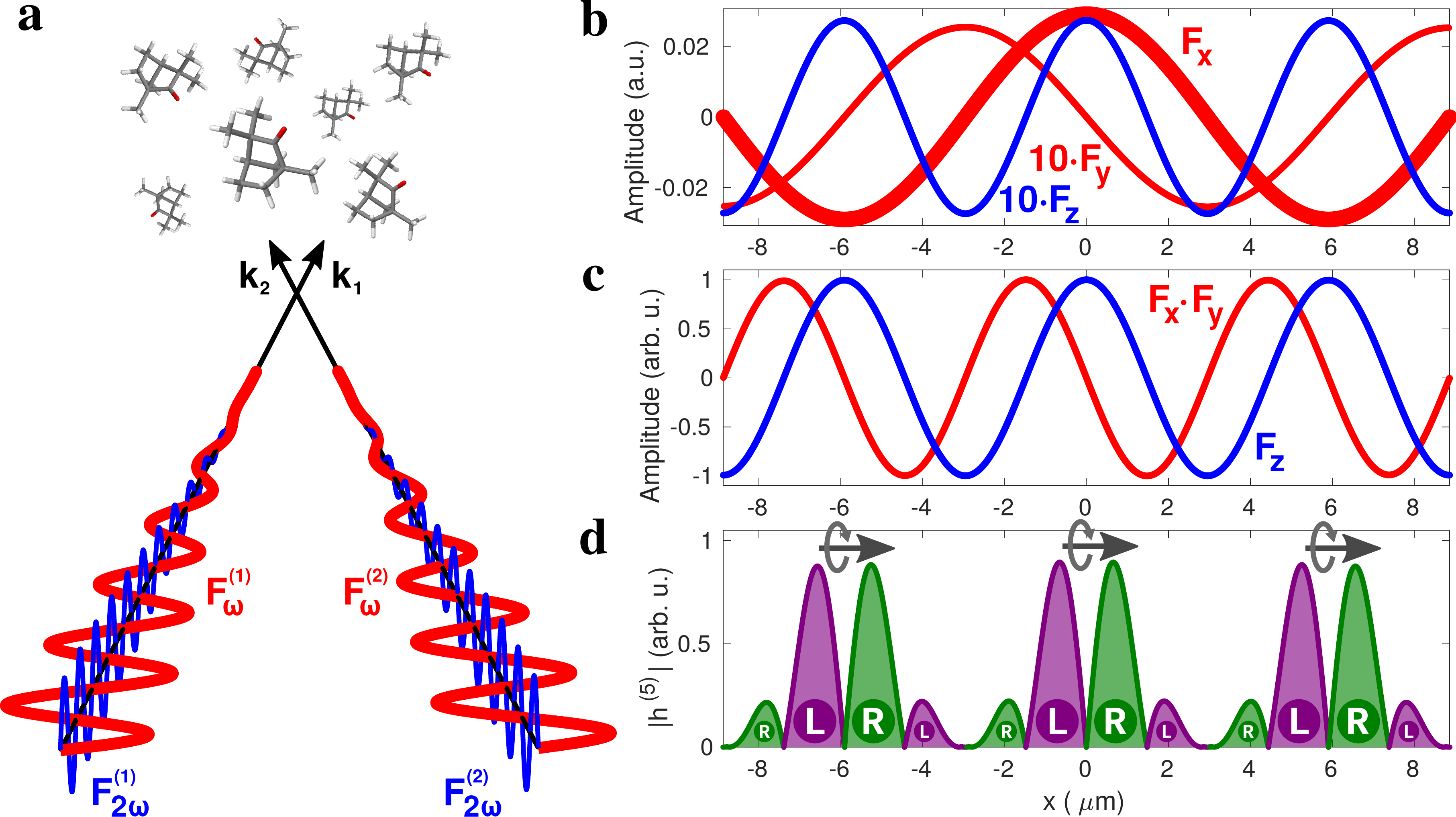}
\caption{\textbf{Racemic space-time ``crystal'' with polarization of handedness.}
\textbf{a,} Setup for creating chirality polarized light. Two non-collinear  beams carry an  $\omega$ field, linearly polarized in the $xy$ propagation plane, and an orthogonally polarized $2\omega$ field, with the same two-color  delay in both beams.
\textbf{b,} Amplitudes of the $x$, $y$ and $z$ field  components  in the overlap region, 
for $\omega=0.044$a.u. ($\lambda=1030$nm),
$F_{\omega}^{(1)}=F_{\omega}^{(2)}= 0.0146$a.u.,
$F_{2\omega}^{(1)}/F_{\omega}^{(1)}=F_{2\omega}^{(2)}/F_{\omega}^{(2)}=\sin{(\alpha)}$;
$2\alpha=10^{\circ}$ is the angle between the beams,
the focal diameter is $200$nm.
\textbf{c,} Normalized $2\omega$-field amplitude ($F_z$) and transverse spin $\propto F_x F_y$.
\textbf{d,} Local handedness of the light field, characterized by its fifth-order chiral correlation function $h^{(5)}$.
The colours encode the phase of $h^{(5)}$ and thus the field's handedness,
which is controlled by the relative phase $\phi$ (see Eq. \ref{eq_Lissajous}); purple: $\arg\{h^{(5)}\}=2\phi+0.5\pi$, green: $\arg\{h^{(5)}\}=2\phi-0.5\pi$.
The grey arrows indicate the direction of polarization of chirality.
}
\label{fig_setup}
\end{figure}

We now illustrate this general analysis with a specific example and demonstrate that racemic, chirality-polarized light can be used to discriminate chiral molecules with extremely high efficiency.
Chirality-polarized light can be created using two beams propagating in the $xy$ plane, at small angles $\alpha=\pm5^{\circ}$ with respect to the $y$ axis, as shown in Fig. \ref{fig_setup}a.
Both contain a fundamental field, linearly polarized in the plane of propagation, and a weak second harmonic component polarized orthogonal to this plane.
In the overlap region, the total electric field is elliptically polarized in the $xy$ plane at frequency $\omega$, and it has a weak, linearly polarized, $2\omega$ frequency component along $z$, 
\begin{equation}\label{eq_Lissajous}
\mathbf{F}(x,t) = \Re\Big\{ \big[ F_x(x)\hat{\mathbf{x}} + iF_y(x)\hat{\mathbf{y}} \big] e^{-i\omega t} +  F_z(x) e^{-2i(\omega t + \phi)}\hat{\mathbf{z}} \Big\}
\end{equation}
where the two-color phase delay $\phi=\frac{\phi_1+\phi_2}{2}$, which controls the field's handedness, is determined by the two-colour phase delays in the individual beams, $\phi_1$ and $\phi_2$.
The spatial modulation of the three orthogonal polarization components, $F_x$, $F_y$ and $F_z$, is presented in Fig. \ref{fig_setup}b.
The electric field vector, at a given point in space, draws a chiral Lissajous figure in  $\mathbf{F}$-space.
Fig. \ref{fig_setup}c shows that 
field's transverse spin
$\mathbf{S}_{2\omega}\propto\mathbf{F}_\omega^*\times\mathbf{F}_\omega\propto F_x F_y\hat{\mathbf{z}}$ and $\mathbf{F}_{2\omega}=F_z\hat{\mathbf{z}}$ change sign at different positions.
As a result, the sign of their product $\mathbf{S}_{2\omega}\cdot\mathbf{F}_{2\omega}$, which determines the sign of light's handedness, changes periodically in space.
This spatial distribution of field's handedness in the near field is recorded in the non-linear response of the medium.   
In the lowest order of non-linearity, the strength of the local enantio-sensitive response is controlled by the chiral correlation  function\cite{Ayuso2019NatPhot} $h^{(5)}\propto \mathbf{S}_{2\omega}\cdot\mathbf{F}_{2\omega}|\mathbf{F}_{\omega}|^2$ shown in Fig. \ref{fig_setup}d.
We see a periodic structure of ``dimers" of alternating handedness, the structure envisioned in Fig. \ref{fig_scheme}b. 
The overall light field  has mirror symmetry with respect to the $yz$ plane (up to a global time shift) and is achiral. However, its handedness is polarized,  with the $x$-component of the dipole of chirality  (Eq. \ref{zilch}) equal to:
\begin{equation}\label{eq_polarization}
 \tilde{h}_x\propto \cos{(\phi_2-\phi_1)} e^{i(\phi_1+\phi_2)} 
\end{equation}
where $\phi_1$ and $\phi_2$ are the two-colour phase delays in each of the  two beams (see Fig. \ref{fig_setup}).
The difference $\phi_2-\phi_1$ controls the amplitude of $\tilde{\mathbf{h}}$, which maximizes for $\phi_1=\phi_2$.
The phase of $\tilde{\mathbf{h}}$ is controlled by $\phi_1+\phi_2$.
This gives us  control over the polarization of the field's handedness:
we set $\phi_1=\phi_2$ to maximize its strength,
and then vary $\phi_1, \phi_2$ synchronously to control the orientation of $\tilde{\mathbf{h}}$ in the complex plane.
Note that the polarization of the field handedness in position space, evident in Fig. \ref{fig_setup}d, leads to a non-zero value of $\int h^{(5)} x dx$, which is proportional to $\tilde{h}_x$ for this definition of the unit cell.

\begin{figure}[H]
\centering
\includegraphics[width=15cm, keepaspectratio=true]{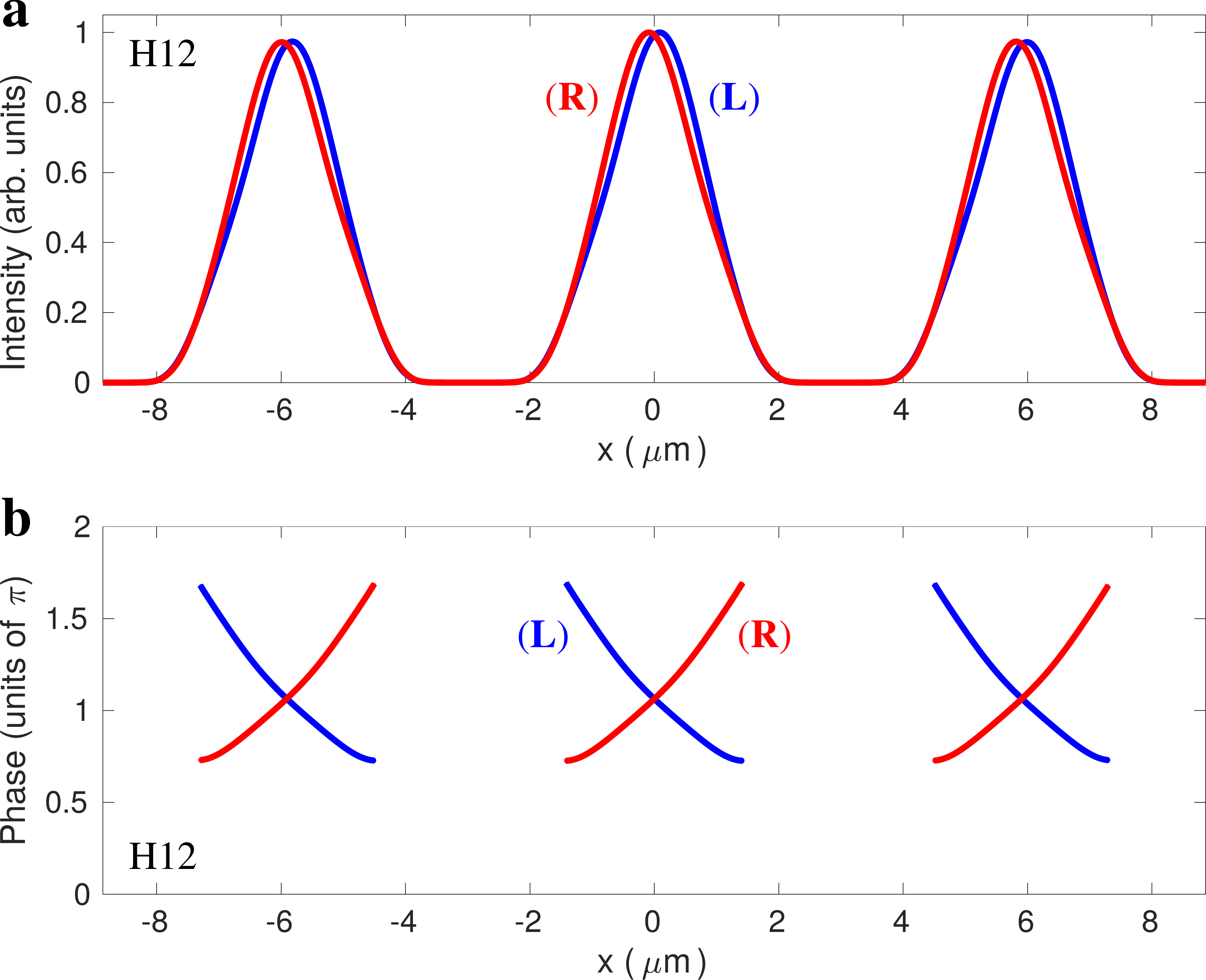}
\caption{\textbf{Enantio-sensitive non-linear response of chiral matter to chirality-polarized light.}
Intensity (\textbf{a}) and phase (\textbf{b}) of the nonlinear response driven by 
the space-time light structure presented in Fig. \ref{fig_setup} in randomly oriented left-handed (blue) and right-handed (orange) fenchone molecules at frequency $12\omega$;
$\lambda=1030$nm,
$F_{\omega}^{(1)}=F_{\omega}^{(2)}= 0.0146$a.u.,
$F_{2\omega}^{(1)}/F_{\omega}^{(1)}=F_{2\omega}^{(2)}/F_{\omega}^{(2)}=0.1$,
$\alpha=5^{\circ}$,
the focal diameter $200$nm.
}
\label{fig_near}
\end{figure}

We now analyze the interaction of the chirality-polarized light field depicted in Fig. \ref{fig_setup} with chiral matter. 
Fig. \ref{fig_near} shows  the
nonlinear response of left- and right-handed randomly oriented fenchone molecules driven by 
the field  in Fig. \ref{fig_setup}. The fundamental wavelength of the incident field 
is $1300$ nm with intensity $7.5\cdot10^{12}$ W$\cdot$cm$^{-2}$ in each beam; the second 
harmonic intensity is $100$ times weaker, and we calculate the response polarized along $z$.
Panels a,b show the intensity and phase of harmonic 12 (for other harmonics, see Supplementary Information).
The response of opposite molecular enantiomers is antisymmetric with respect to $x$; the effect of exchanging the molecular enantiomer is equivalent to reversing the polarization of chirality of the field, 
which can be done by shifting the two-colour phase delay in both beams (see Eq. \ref{eq_polarization}).

The single-molecule response, at a given point space, is enantio-sensitive in intensity (Fig. \ref{fig_near}a) because the driving field is locally chiral.
However, the overall light field is achiral, and thus the total intensity signal, obtained after integration over  $x$, is identical in left- and right-handed molecules.
Still, the direction of polarization of the field's handedness is imprinted in the phase of the nonlinear response, which depends strongly on the molecular handedness (Fig. \ref{fig_near}b):
the slope of the phase dependence on $x$ is positive for right-handed molecules and negative in left-handed molecules.
These enantio-sensitive phase gratings control the direction of macroscopic harmonic emission in the far field.
The strength of this effect is controlled by the dipole of chirality $\tilde{{h}}_x$, which determines the deflection angle.

\begin{figure}[H]
\centering
\includegraphics[width=\linewidth, keepaspectratio=true]{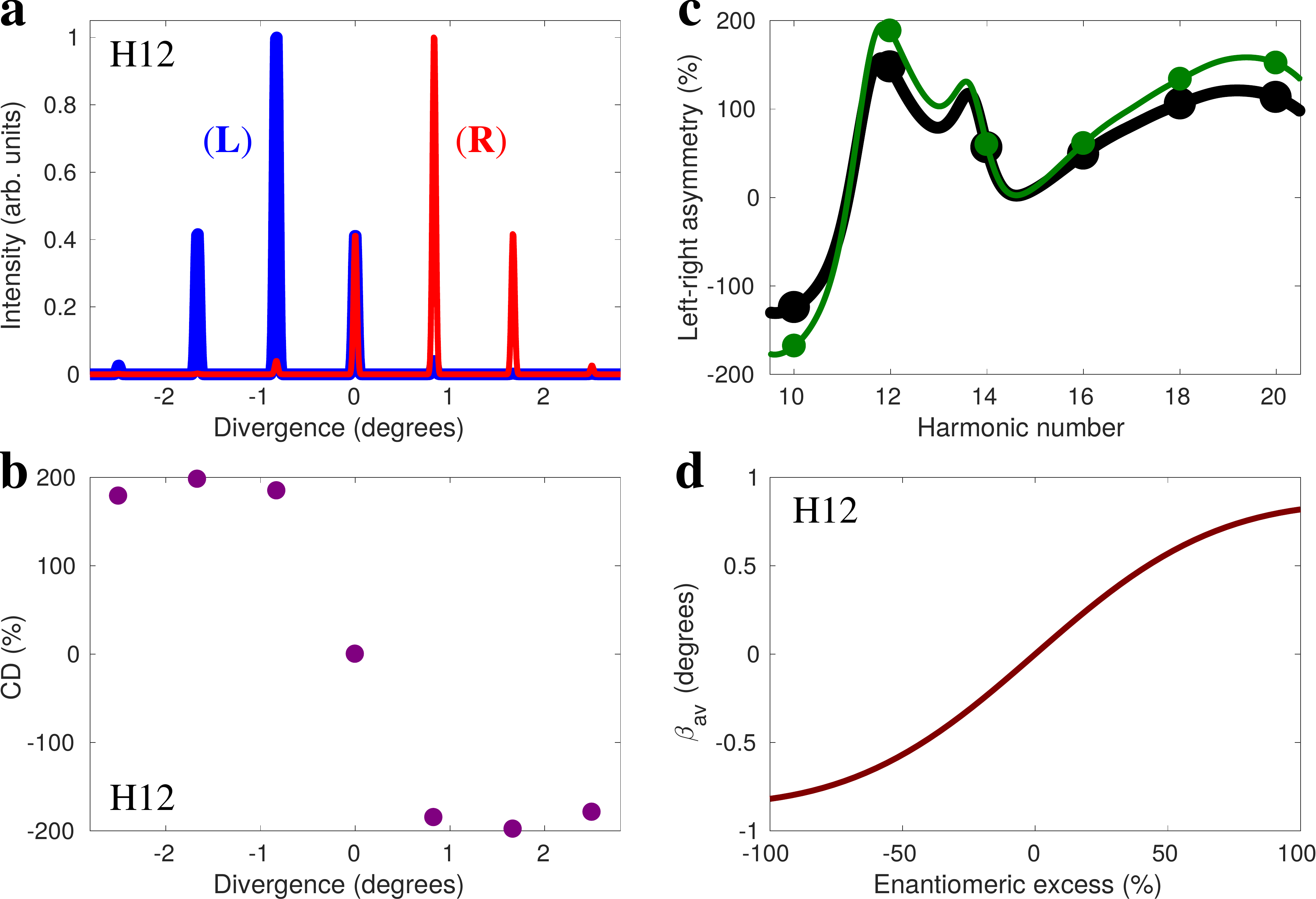}
\caption{\textbf{Macroscopic high harmonic generation.}
\textbf{a,} Macroscopic high harmonic emission at frequency $12\omega$ from left-handed (blue) and right-handed (red) randomly oriented fenchone molecules as a function of the emission angle.
Field parameters are the same as in Fig.3.
\textbf{b,} Chiral dichroism, $CD=2\frac{I_L-I_R}{I_L+I_R}$.
\textbf{c,} Left-right asymmetry in the macroscopic emission of even harmonics from 10 to 20, calculated including (black) or not including (green) the central emission peak.
\textbf{d,} Mean value of the emission angle as a function of the enantiomeric excess.
}
\label{fig_far}
\end{figure}

Fig. \ref{fig_far}a shows the harmonic $12$ intensity in the far field.
The total (angle-integrated) intensity is the same for left- and right-handed molecules, as in the near field (Fig. \ref{fig_near}a).
However, the direction of emission is extremely enantio-sensitive:
while the left-handed molecules emit harmonics to the left  (towards negative $x$), the right-handed molecules emit harmonics to the right (positive $x$).
We control the enantio-sensitive direction of emission by controlling the polarization of the field's handedness in our setup (see Eq. \ref{eq_polarization}).
Fig. \ref{fig_far}b shows that chiral dichroism resolved in the emission angle
reaches the ultimate efficiency limit, $CD=200\%$ (Fig. \ref{fig_far}b).
We find giant enantio-sensitivity in the direction of emission of all even harmonics, see Supplementary Information.

We can define the left-right asymmetry in the harmonic emission as $A_{LR}=2\frac{I_L-I_R}{I_L-I_R}$, where $I_L$ and $I_R$ are the intensities of harmonic emission to the left ($\beta<0$) and to the right ($\beta>0$), respectively.
This angle-integrated quantity reaches very high values for all harmonic numbers, as shown in Fig. \ref{fig_far}c.
The direction of harmonic emission is correlated to the enantiomeric excess of the medium $ee=\frac{C_R-C_S}{C_R+C_S}$, with $C_R$ and $C_S$ being the concentrations of the right- and left-handed molecules, see Fig. \ref{fig_far}d.
The expectation value of the emission angle is then given by
\begin{equation}\label{eq_beta_ee}
\langle\beta\rangle = \frac{ee \tilde{I}_{aR}}
                               {\tilde{I}_a^{\beta} + ee^2 \tilde{I}_R^{\beta}}
\end{equation}
where $\tilde{I}_{aR}$, $\tilde{I}_a^{\beta}$ and $\tilde{I}_R^{\beta}$ are angle-integrated quantities that do not depend on $ee$.
Eq. \ref{eq_beta_ee} allows us to quantify small values of enantiomeric excess in macroscopic mixtures.

Polarization of chirality is a powerful concept which allows one to engineer highly efficient chiral interactions of racemic objects. It opens several opportunities. 
The first  stems from flexibility in chiral polarization shaping of synthetic chiral light, where 
the spatial dependence of the field's transverse spin $\mathbf{S}^{(n)}_{\omega_n}(\rr)$ and the electric field component $\mathbf{F}_{\omega_n}(\rr)$ parallel to it can be controlled separately. Here one can take full advantage of modern light shaping techniques including polarization shaping in space and time\cite{brixner2001femtosecond,garg2018ultimate,hernandez2017extreme}. In contrast, in standard circularly polarized light, this opportunity is limited, since its electric and magnetic components are locked to each other.
A non-zero dipole of  chirality  is present in any locally chiral field\cite{Neufeld2019PRX,Ayuso2019NatPhot} where field's transverse spin  $\mathbf{S}^{(n)}_{\omega_n}(\rr)$ and   $\mathbf{F}_{\omega_n}(\rr)$  have opposite parity. 

The second  opportunity is to  use chirality polarized light to imprint polarization of chirality onto racemic matter, creating "chiral polaritons". The phase relationship between the imparted angular momentum and polarization waves, induced in the medium by chirality-polarized light, will define the medium chiral polarization, its coherence length, and the strength of its interaction with other chiral structures.

The third opportunity is to exploit giant enantio-sensitivity to 
not only create chirality-polarized matter, but also to control it on ultrafast time scales. 
Chirality-polarized light may  allow us to identify racemic aggregates of chiral 
matter that exhibit complex chirality patterns in space, and to quantify their 
degree of polarization of chirality.

Finally,   polarization of chirality  can also be used for efficient 
separation of opposite enantiomers in racemic mixtures by extending the  
proposal of Ref.\cite{Cameron2014JPCA} to  chirality polarized light, which would allow one to bypass the use mechanical transition gratings and weak  magnetic field interactions.

\section*{Acknowledgements}
DA, AO, MI and OS acknowledge support from the DFG SPP 1840 ``Quantum Dynamics in Tailored Intense Fields'' and DFG grant SM 292/5-1.
MI acknowledges MURI grant EP/N018680/1.

\section*{Competing Interests}
The authors declare that they have no competing financial interests.
The data that support the plots within this paper and other findings of this study are available from the corresponding authors upon reasonable request. 
Correspondence should be addressed to david.ayuso@mbi-berlin.de, andres.ordonez@mbi-berlin.de and olga.smirnova@mbi-berlin.de. 

\newpage

\bibliography{Bibliography}

\end{document}